\documentclass[prd,twocolumn,showpacs,nofootinbib,amsmath,amssymb,floatfix,superscriptaddress,showkeys]{revtex4}
\usepackage{multirow}
\usepackage{graphicx}
\usepackage{epstopdf}
\usepackage{dcolumn}
\usepackage{bm}
\usepackage{threeparttable}
\usepackage{subfigure}
\usepackage{color,txfonts}
\usepackage{ulem}
\definecolor{blue0}{rgb}{0,0,0.6}
\usepackage[colorlinks,linkcolor=blue0,anchorcolor=blue0,citecolor=blue0,urlcolor=blue0]{hyperref}

\newcommand{\jcap}{J. Cosmol. Astropart. Phys.}

\newcommand{\mnras}{Mon. Not. R. Astron. Soc.}

\newcommand{\aap}{Astron. Astrophys}
\newcommand{\aj}{Astron. J.}

\newcommand{\apjl}{Astrophys. J.}

\newcommand{\beq}{\begin{equation}}
\newcommand{\eeq}{\end{equation}}
\newcommand{\beqa}{\begin{eqnarray}}
\newcommand{\eeqa}{\end{eqnarray}}

\begin{document}

\title{Searching for dark-matter induced neutrino signals in dwarf spheroidal galaxies using 10 years of IceCube public data}
\author{Xue-Kang Guo}
\author{Yi-Fei L{\"u}}
\author{Yong-Bo Huang}
\email[]{huangyb@gxu.edu.cn}
\author{Rong-Lan Li}
\author{Ben-Yang Zhu}
\author{Yun-Feng Liang}
\affiliation{Guangxi Key Laboratory for Relativistic Astrophysics, School of Physical Science and Technology, Guangxi University, Nanning 530004, China}

\date{\today}

\begin{abstract}
This study searches for neutrino signals from 18 dwarf spheroidal galaxies (dSphs) using 10 years of publicly available muon-track data of the IceCube neutrino observatory. We apply an unbinned likelihood analysis on each of these dSphs to derive the significance the putative neutrino emission. To further enhance our sensitivity, we also stack all dSphs together to perform a joint analysis. However, no significant neutrino emission signal was detected in either the single-source or stacking analysis. Based on these null results, we derive constraints on the annihilation cross section of dark matter particles. Compared to the existing literature, our constraints via the channel $\chi\chi\rightarrow\mu^+\mu^-$ are comparable to the ones from the VERITAS observations of dSphs.
\end{abstract}

\maketitle

\section{Introduction}
\label{sec:intro}

In the past few decades, researchers are inclined to agree with the existence of dark matter (DM) as observational evidence accumulates, including but not limited to the Bullet cluster, strong gravitational lensing and cosmic microwave background radiation \cite{2017FrPhy..12l1201Y}. Despite the nature of dark matter remains a mystery, people can infer what it is from its properties that we have known, e.g., widely recognized as electric and chromatic neutrality, gravitationally and (sub)weakly interacting \cite{2018RPPh...81f6201R}. A popular inference claims that the dominant fraction of DM is probably ‘cold’, in other words it should be non-relativistic and massive. Many models of dark matter have been proposed, such as weakly interacting massive particles (WIMPs), axion-like particles (ALPs), and right-handed (or “sterile”) neutrinos \cite{2017FrPhy..12l1201Y}. 

WIMP has been considered as the most hopeful dark matter candidate in the past tens of years. However, research on WIMP has encountered a bottleneck state since no reliable signal has been observed in the direct or indirect dark matter detection experiments. The strongest limits, placed by XENON1T, LUX, and PandaX-II \cite{2019JPhG...46j3003S}, only lies a few orders of magnitude above the “neutrino floor” \cite{2022EPJC...82..650N,2014PhRvD..89b3524B}, which means most of the WIMP theories might be excluded. It seems not optimistic to find a convincing signal of WIMP in the rest narrow region, which has motivated researchers to explore the dark matter physics in a wider mass region. 

Ultra-heavy dark matter (UHDM) has attracted many researchers' attention in recent years as it presents an alternative mass range for dark matter, roughly from 10 TeV to the Plank energy ($\sim10^{19}$ GeV) \cite{2023ApJ...945..101A}. Many works have been carried out to search for the decaying UHDM \cite{Cohen:2016uyg,Kalashev:2016cre,IceCube:2018tkk,Kachelriess:2018rty,Bhattacharya:2019ucd,Ishiwata:2019aet,Chianese:2021jke,Maity:2021umk,IceCube:2022vtr,lhaaso22decay}. For the annihilation UHDM, there exists a general upper bound imposed by unitarity on the particle mass of thermal relic dark matter \cite{2017FrPhy..12l1201Y}. DM particles are widely believed to originate in thermal equilibrium with equal rates of production and annihilation in the early universe. As the universe expands and cools, DM particles cool down and become nonrelativistic and the equilibrium is broken. The production of dark matter becomes slower than annihilation, until the expansion eventually shuts this process off and the relic abundance freezes out. The thermal production mechanism dictates that the cross section cannot be arbitrarily large, which translates into the upper bound on the DM mass \cite{2022ApJ...938L...4T}. The limit has been lowered to 144 TeV by \cite{2019PhRvD.100d3029S}. Though this unitarity bound has long discouraged people from searching for annihilation UHDM, several mechanisms have been proposed to violate this bound \cite{2022arXiv220306508C}. 

Neutrinos, though not a good DM candidate (because its thermal decoupling temperature is too high ($\sim$1 MeV) for a cold DM candidate), serve as a new medium for indirect detection of dark matter in multi-messenger era \cite{2019ICRC...36...16W}. Having essentially tiny masses and no electric charge, neutrinos travel through the universe in an unattenuated and undeviated way, making them an ideal astronomical messenger \cite{2018PrPNP.102...73A}. The extremely weak interactions of neutrinos with matter pose a significant challenge for neutrino observation. However, with the completion of IceCube in 2011, the situation has remarkably improved. This study utilizes 10 years of IceCube muon-track data (April 2008 to July 2018) \cite{2021arXiv210109836I,2008APh....29..299B} and an unbinned maximum-likelihood-ratio method, which is commonly used in previous research \cite{2010PhRvD..82l3503E,2010PhRvD..81h3506S,2023arXiv230313663T,2021PhRvD.103l3018Z}, to search for DM induced neutrinos from dwarf spheroidal galaxies (dSphs). More details about IceCube will be provided in the second section.

DSphs are always seen as the optimal targets for indirect dark matter searches, since they are DM-rich regions with negligible astrophysical backgrounds and short distances to the Earth ($<0.5$ Mpc)\cite{2004PhRvD..69l3501E}. If there are high-energy neutrinos found in the direction of a dSph, they are probably produced by the annihilation of dark matter, since it is not expected any other astrophysical mechanism could generate high-energy neutrinos at TeV energies in dSphs. Currently, more than 60 dwarf satellite galaxies or candidates have been discovered by wide-field optical imaging surveys, including Sloan Digital Sky Survey (SDSS) \cite{2000AJ....120.1579Y,2007ApJ...654..897B}, Gaia \cite{2018A&A...616A...1G}, DES \cite{2015ApJ...807...50B,2015ApJ...805..130K}, etc. In this work, we choose 18 dSphs as our targets to search for UHDM signals, both “classical” and “ultra-faint” dSphs are included. As IceCube’s muon-track strategy is most sensitive to the sources from the northern sky\footnote{For the neutrinos from the northern half of the sky, the Earth is used as a filter to remove the huge background of atmospheric muons \cite{2018PrPNP.102...73A}.}, dSphs located in the southern sky are all excluded. It should be noted that Triangulum II was excluded for its anomalously high J-factor and large uncertainty. 
The adopted dSphs and their J-factors are listed in Table~\ref{tab:tab1}. The J-factors are extracted from \cite{2018ApJ...860...66M,2016JCAP...09..047H}.

\begin{table}[h]
\renewcommand\arraystretch{1.2}
\caption{Information of the 18 dSphs considered in this work.}
\label{tab:tab1}
\centering
\begin{tabular}{lcccc}
\hline
\hline
DSph & $\begin{array}{c}\text{RA} \\ (\mathrm{deg})\end{array}$ & $\begin{array}{c}\text{Dec} \\ (\mathrm{deg})\end{array}$ & $\begin{array}{c}\text{Distance} \\ (\mathrm{kpc})\end{array}$ & $\begin{array}{c}\log_{10}J \\ (\mathrm{GeV}^{2}\mathrm{~cm}^{-5})\end{array}$ \\
\hline
\textit{Boötes I} & 210.02 & 14.51 & 66 & $18.5_{-0.4}^{+0.7}$ \\
\textit{Canes Venatici I} & 202.01 & 33.55 & 218 & $17.5_{-0.2}^{+0.4}$ \\
\textit{Canes Venatici II} & 194.29 & 34.32 & 160 & $18.5 _{-0.9}^{+1.2}$ \\
\textit{Coma Berenices} & 186.75 & 23.91 & 44 & $19.6 _{-0.7}^{+0.8}$ \\
Draco & 260.07 & 57.92 & 82 & $19.1 _{ -0.3}^{+0.4}$ \\
\textit{Draco II} & 238.17 & 64.58 & 20 & $18.1 _ {-3.2}^{+2.4}$ \\
\textit{Hercules} & 247.77 & 12.79 & 132 & $17.5 _ {-0.7}^{+0.7}$ \\
Leo I & 152.11 & 12.31 & 250 & $17.8 _{ -0.2}^{+0.5}$ \\
Leo II & 168.36 & 22.15 & 205 & $18.0 _{-0.2}^{+0.6}$ \\
\textit{Leo V} & 143.72 & 17.05 & 180 & $16.1 _{ -1.1}^{+1.2}$ \\
\textit{Leo T }& 172.79 & 2.22 & 407 & $17.6 _{ -0.5}^{+1.0}$ \\
\textit{Pisces II} & 344.63 & 5.95 & 182 & $16.9 _{ -1.7}^{+1.5}$ \\
\textit{Segue 1} & 151.75 & 16.08 & 23 & $17.2 _{ -2.3}^{+1.8}$ \\
\textit{Segue 2} & 34.82 & 20.16 & 35 & $18.9 _{-0.9}^{+1.1}$ \\
\textit{Ursa Major I} & 158.77 & 51.95 & 97 & $18.7 _{-0.4}^{+1.6}$ \\
\textit{Ursa Major II} & 132.87 & 63.13 & 30 & $19.9 _{-0.6}^{+0.6}$ \\
Ursa Minor & 227.24 & 67.22 & 66 & $19.0 _{-0.1}^{+0.1}$ \\
\textit{Willman 1} & 162.34 & 51.05 & 38 & $19.5 _{-0.6}^{+1.2}$ \\
\hline
\hline
\multicolumn{5}{{p{.47\textwidth}}}{Note: The distances and J-factors of the dSphs are taken from \cite{2016JCAP...09..047H} (please see the references therein for the original sources). Ultrafaint dSphs are specially marked in italics. Notice that the kinematic analysis of Segue 1 is extremely sensitive to the member stars selection, bringing this target a large uncertainty. A larger median value of J-factor of this source is also claimed \cite{2018ApJ...860...66M}.}
\end{tabular}
\end{table}

\section{IceCube observations}
The IceCube Neutrino Observatory is a kilometer-scale neutrino detector buried under the Antarctic Ice Sheet \cite{IceCube:2016zyt}. The instrument consists of 86 vertical strings at a depth of 1450 m and each string is equipped with 60 digital optical modules (DOMs). When a high-energy neutrino interacts with an atomic nucleus in or near the instrumented volume, secondary charged particles will be produced at superluminal speed moving through the ice and creating Cherenkov radiation, which could be detected by the DOMs and converted to electrical signals. The neutrinos of three flavors correspond to two basic kinds of events, namely, the muon neutrinos trigger track events and the electron/tau neutrinos trigger electromagnetic or hadronic shower events which look like a round or blob. The track events have a better angular resolution (as good as $<1^\circ$) but worse energy resolution ($\sim$200\% at $\sim$100 TeV) compared to the shower events ($\sim10^\circ-15^\circ$ angular resolution and $\sim$15\% energy resolution above 100 TeV)  \cite{2015ICRC...34.1109N}. 

IceCube is sensitive to the primary neutrinos from 100 TeV to tens of EeV. The wide energy reach provides IceCube a particular advantage in indirect searches for high-mass DM \cite{2022arXiv220306508C,2016NPPP..279...47G,IceCube:2011dm,IceCube:2013bas,IceCube:2015rnn,IceCube:2016aga,IceCube:2017rdn,IceCube:2018tkk,ANTARES:2020leh,IceCube:2021xzo,IceCube:2022clp,IceCube:2023ies,Murase:2015gea,Chianese:2017nwe,Kachelriess:2018rty}. In this work, we use 10 years of muon-track data released by the IceCube Collaboration \cite{IceCube:2021xar}. In total, 1134450 muon-track events are included, and grouped into ten samples, denoted as (i) IC40, (ii) IC59, (iii) IC79, and (iv-x) IC86-I–IC86-VII. Each sample corresponds to a single season of IceCube data-taking, including roughly one year of data. The numbers in the names represent the numbers of strings in the detector\cite{2021PhRvD.103l3018Z}. 

The released data are composed of four parts: the experimental data events, detector uptime, instrument smearing matrices and effective areas \cite{2021arXiv210109836I}. The data event list contains the reconstructed information of particles from the IceCube's point source neutrino selection, including MJD time, reconstructed energy and direction of muon and the estimated angular uncertainty. The detector uptime gives information on the time intervals when IceCube is in good runs. The instrument smearing matrices and effective areas model the response of the IceCube detector to neutrino signals.

\section{data analysis}
\label{sec:data-analysis}

We search for neutrino emission from dSphs using an unbinned maximum-likelihood method. If no signal is found, the upper limit (UL) at a 95\% confidence level (C.L.) on the neutrino flux is derived. By requiring the model-expected neutrino spectrum of dark matter annihilation from dSphs not to exceed the flux upper limit, the constraint on the annihilation cross-section can be obtained. In our analysis, dSphs will be seen as point sources, since their extension is negligible compared to the angular uncertainty on the reconstructed direction of IceCube events.

\subsection{likelihood function}
Our analysis uses an unbinned maximum-likelihood ratio method which is commonly used in previous works \cite{Braun:2008bg,2018PrPNP.102...73A,Hooper:2018wyk,Smith:2020oac,2021PhRvD.103l3018Z,Chang:2022hqj,2022PhRvD.106h3024L}. The likelihood function is given by the product of probability density functions (PDFs) of each muon-track event (indexed by $i$ ) in the ten data samples (indexed by $k$ ):
\begin{equation}
L\left(n_{s}\right)=\prod_{k} \prod_{i \in k}\left[\frac{n_{s}^{k}}{N_{k}} S_{i}^{k}+\left(1-\frac{n_{s}^{k}}{N_{k}}\right) B_{i}^{k}\right],
\label{eq:like}
\end{equation}
where $n_{s}^{k}$ is the number of signal events from the sources in the sample $k$, and $N_{k}$ is the total number of events within region of interest (ROI) and in the sample $k$. The ROI is defined as a $5^{\circ}$ circle region around the target source.

The $S_{i}^{k}$ and $B_{i}^{k}$ in Eq.~(\ref{eq:like}) are the signal and background PDFs. We consider that the events from a source direction $\vec{x}_{s}$ should follow a 2-Dimension Gaussian distribution in the reconstructed direction $\vec{x}_{i}$,
\begin{equation}
S_{i}^{k}=S_{\mathrm{spat}}^{k}\left(\vec{x}_{i} \mid \sigma_{i}, \vec{x}_{s}\right)=\frac{1}{2 \pi \sigma_{i}^{2}} \exp \left(-\frac{D\left(\vec{x}_{i}, \vec{x}_{s}\right)^{2}}{2 \sigma_{i}^{2}}\right)
\label{eq:si1}
\end{equation}
where $D\left(\vec{x}_{i}, \vec{x}_{s}\right)$ represents the angular distance between the true and reconstructed directions, and $\sigma_{i}$ represents the uncertainty of the reconstructed direction of the event. While the background PDF
\begin{equation}
B_{i}^{k}=B_{\mathrm{spat}}^{k}\left(\delta_{i}\right)=\frac{N_{\delta_{i} \pm 3}^{k}}{N_{k} \times \Delta \Omega}
\label{eq:bi1}
\end{equation}
is obtained from the event distribution of the signal-free region \cite{2021PhRvD.103l3018Z}. The $N_{\delta_{i} \pm 3}^{k}$ here is the number of events within a ring region of $\delta_{i} \pm 3^{\circ}$ and $\Delta \Omega$ is its solid angle. Since IceCube is located at the geographic South Pole, $B_{i}$ is only declination dependent.

The test statistic (TS) is defined as
\begin{equation}
\mathrm{TS}=2 \ln \frac{L\left(\hat{n}_{s}\right)}{L\left(n_{s}=0\right)}
\end{equation}
The denominator in the above equation refers to the null hypothesis that all the events come from the background. The best-fit number of signal events $\hat{n}_{S}$ is obtained by maximizing the likelihood value when $n_{s}$ is free to vary.

\subsection{energy term in the signal/background PDFs}
In many previous works of analyzing IceCube muon data, the energy PDF term was ignored. As is suggested in Ref~\cite{Braun:2008bg}, the addition of the energy PDF term is expected to improve the sensitivity by a factor of $\sim 2$. In this work, we include the energy term into the signal and background PDFs, which now consist of a spatial part and an energy part:
\begin{equation}
S_{i}^{k}=S_\mathrm{spat}^{k}\left(\vec{x}_{i} \mid \sigma_{i}, \vec{x}_{s}\right) \times S_{\mathrm{ener}}^{k}\left(E_{i} \mid \vec{x}_{s}, \gamma\right),
\label{eq:si2}
\end{equation}
\begin{equation}
B_{i}^{k}=B_\mathrm{spat}^{k}\left(\delta_{i}\right) \times B_{\mathrm{ener}}^{k}\left(E_{\mathrm{rec}} \mid \delta_{i}\right)
\label{eq:bi2}
\end{equation}
where $S_{\mathrm{spat}}^{k}$ and $B_{\mathrm{spat}}^{k}$ are just the ones of Eq.~(\ref{eq:si1}) and Eq.~(\ref{eq:bi1}).

For the energy term of signal PDF, it is \cite{2022MNRAS.514..852H} :
\begin{equation}
S_{\mathrm{ener}}^{k}\left(E_{i} \mid \vec{x}_{s}, \gamma\right)=\frac{\int \Phi_{v}\left(E_{v}\right) A_{\mathrm{eff}}^{k}\left(E_{v}, \delta_{s}\right) M_{k}\left(E_{i} \mid E_{v}, \delta_{s}\right) d E_{v}}{\int \Phi_{v}\left(E_{v}\right) A_{\mathrm{eff}}^{k}\left(E_{v}, \delta_{s}\right) d E_{v}}
\end{equation}
with $A_{\rm eff}^{k}$ the effective area and $M_{k}$ the instrument smearing matrices. For the background PDF, the energy part is given by:
\begin{equation}
B_{\mathrm{ener}}^{k}\left(E_{\mathrm{rec}} \mid \delta_{i}\right)=\frac{N_{i j}^{k}}{N_{\delta_{i} \pm 3}^{k} \Delta E_{j}}
\end{equation}
where $N_{i j}^{k}$ is the number of events within the declination range of $\delta \in\left[\delta_{i}-3, \delta_{i}+3\right)$ and the reconstructed energy range of $\log _{10} E_{\mathrm{rec}} \in\left[\log _{10} E_{j}, \log _{10} E_{j}+0.1\right)$ for the sample $k$.

In Eq.~(\ref{eq:like}), the signal counts in the sample $k$, $n_{s}^{k}$, is given by
\begin{equation}
n_{s}^{k}=t_{k} \int A_{\mathrm{eff}}^{k}\left(E_{v}, \delta\right) \Phi_{v}\left(E_{v}\right) d E_{v}
\end{equation}
In this work, the spectral model $\Phi_{v}\left(E_{v}\right)$ would be a single power-low spectrum or a spectrum of DM annihilation (see Sec.~\ref{sec:C}). As a baseline spectral model, the single power-law spectrum is:
\begin{equation}
\Phi_{v}\left(E_{v}\right)=\Phi_{0} \times\left(\frac{E_{v}}{100\,\mathrm{TeV}}\right)^{-\gamma}
\end{equation}
where $\Phi_{0}$ is the neutrino flux at $100\,\mathrm{TeV}$. For neutrinos of astrophysical origin, the spectrum index $\gamma$ in the range from 2.0 to 3.0 is usually expected. For example, the most recent work gives the updated best-fit $\gamma=2.28_{-0.09}^{+0.08}$ for the IceCube's diffuse neutrinos based on the measurements using 10 years of muon-track events between a few $\mathrm{TeV}$ and 10 $\mathrm{PeV}$ \cite{2019ICRC...36.1017S}, while the value increases to $2.37^{+0.08}_{-0.23}$ after combing the cascade analysis \cite{2022icrc.confE1129G}. These results are in a good agreement with the expectation from the Fermi shock acceleration mechanism \cite{2020JCAP...09..007D}. Thus (when no neutrino emission is detected) we fix the spectral index to 2.3 in the case of using the PL spectrum.

\subsection{DM annihilation flux}
\label{sec:C}
The differential neutrino flux from dark matter annihilation can be written as:
\begin{equation}
\Phi_{v}\left(E_{v}\right)=\frac{\langle\sigma v\rangle}{8 \pi m_{\chi}^{2}} \frac{dN_{v}}{dE_{v}}\left(E_{v}\right) J_{\mathrm{ann}}
\label{eq:flux}
\end{equation}
where $\langle\sigma v\rangle$ is the velocity-averaged DM self-annihilation cross section, $m_{\chi}$ is the DM particle mass. The $d N_{v} / d E_{v}$ is the differential neutrino yields per dark matter annihilation \cite{2023arXiv230313663T}, which we calculate using HDMSpectra, a Python package providing tabulated dark matter decay and annihilation spectra for dark matter masses between the $\mathrm{TeV}$ and Planck scale for various annihilation channels \cite{2021JHEP...06..121B}. We consider 6 channels: $\chi\chi\rightarrow t\bar{t}, b\bar{b}, \tau^{+}\tau^{-}, \mu^{+}\mu^{-}, W^{+}W^{-}$ and $v_{\mu}\bar{v}_{\mu}$, corresponding to the typical channels of quarks, leptons and bosons. According to the energy distributions between different final-states particles shown in \cite{2011JCAP...03..051C}, these channels have relatively higher neutrino yields. In realistic models, the differential yield per annihilation $dN_{v}/dE_{v}$ in Eq.~(\ref{eq:flux}) is a sum over different final states: $dN_v/dE_v=\sum_{f}B_{f}dN_{v}^{f}/dE_{v}$, where $B_{f}$ is the branching fraction into final state $f$. In this work, we consider the DM annihilate to a single final state, i.e. the branching fraction is assumed to be 100\% for each of the 6 channels. In addition, considering neutrino oscillations, the $dN_{v}/dE_{v}$ should be contributed by neutrinos of three flavors, 
\begin{equation}
    \frac{dN_{v}}{dE_{v}}=\frac{1}{3}\times\left(\frac{dN_{v_e}}{dE_{v_e}}+\frac{dN_{v_\mu}}{dE_{v_\mu}}+\frac{dN_{v_\tau}}{dE_{v_\tau}}\right).
\end{equation}

The $J_{\mathrm{ann}}$ in Eq.~(\ref{eq:flux}) represents the astrophysical J-factor, which is the square of the dark matter density integrated along the line of sight (l.o.s) and over the region of interest (ROI),
\begin{equation}
J_{\mathrm{ann}}=\int_{\mathrm{ROI}}d\Omega\int_{\mathrm{l.o.s}} \rho^{2}(r(l,\theta))dl
\end{equation}
where $\Omega$ denotes the solid angle of the ROI and the radius from the dSph center is $r(l,\theta)=\sqrt{D^{2}-2lD\cos\theta+l^{2}}$ with $D$ the distance of the dSph. In our work, we adopt the J-factors reported in Ref~\cite{2018ApJ...860...66M,2016JCAP...09..047H}. In the calculation of these J-factors, the dark matter density profile, $\rho({r})$, adopts the Einasto profile \cite{2015ApJ...801...74G}:
\begin{equation}
\rho(r)=\rho_{-2} \exp \left\{-\frac{2}{\alpha}\left[\left(\frac{r}{r_{-2}}\right)^{\alpha}-1\right]\right\}
\end{equation}
This profile introduces an extra shape parameter $\alpha$ with respect to the standard Navarro-Frenk-White (NFW) profile \cite{1997ApJ...490..493N}. The median values of J-factors and their statistical uncertainties of 18 $\mathrm{dSphs}$ are listed in Table~\ref{tab:tab1}.

\begin{table}[]
\caption{Analysis results for the 18 dSph galaxies.}
\label{tab:tab2}
\centering
\begin{tabular}{p{2.7cm}p{1.3cm}p{1cm}c}
\hline
\hline
dSph & $\hat{n}_{s}$ & TS & $\begin{array}{c}\Phi_{\mathrm{UL}} @ 100\,\mathrm{TeV} \\ {(10^{-19}\,\mathrm{cm^{-2}s^{-1}GeV^{-1}})}\end{array}$ \\
\hline
\textit{Boötes I} & 0.0 & 0.0 & 0.70 \\
\textit{Canes Venatici I} & 3.3 & 0.7 & 0.74 \\
\textit{Canes Venatici II} & 0.0 & 0.0 & 0.32 \\
\textit{Coma Berenices} & 1.7 & 0.1 & 0.61 \\
Draco & 2.8 & 0.1 & 1.05 \\
\textit{Draco II} & 0.0 & 0.0 & 0.61 \\
\textit{Hercules} & 11.3 & 2.6 & 1.27 \\
Leo I & 2.2 & 0.1 & 0.88 \\
Leo II & 0.0 & 0.0 & 0.32 \\
\textit{Leo V} & 0.3 & 0.0 & 0.78 \\
\textit{Leo T} & 1.8 & 0.1 & 0.73 \\
\textit{Pisces II} & 0.0 & 0.0 & 0.34 \\
\textit{Segue 1} & 5.6 & 0.6 & 1.01 \\
\textit{Segue 2} & 0.0 & 0.0 & 0.26 \\
\textit{Ursa Major I} & 9.7 & 1.0 & 1.43 \\
\textit{Ursa Major II} & 1.1 & 0.1 & 0.68 \\
Ursa Minor & 0.0 & 0.0 & 0.42 \\
\textit{Willman 1} & 0.0 & 0.0 & 0.69 \\
\hline
\hline
\multicolumn{4}{{p{.45\textwidth}}}{Note: Columns 2 and 3 show the best-fit number of signal events and TS value of the putative neutrino emission in the direction of each dSph. Column 4 shows the upper limit on the neutrino flux at $100\,\mathrm{TeV}$.}
\end{tabular}
\end{table}

\begin{figure}
    \centering
    \includegraphics[width=0.9\columnwidth]{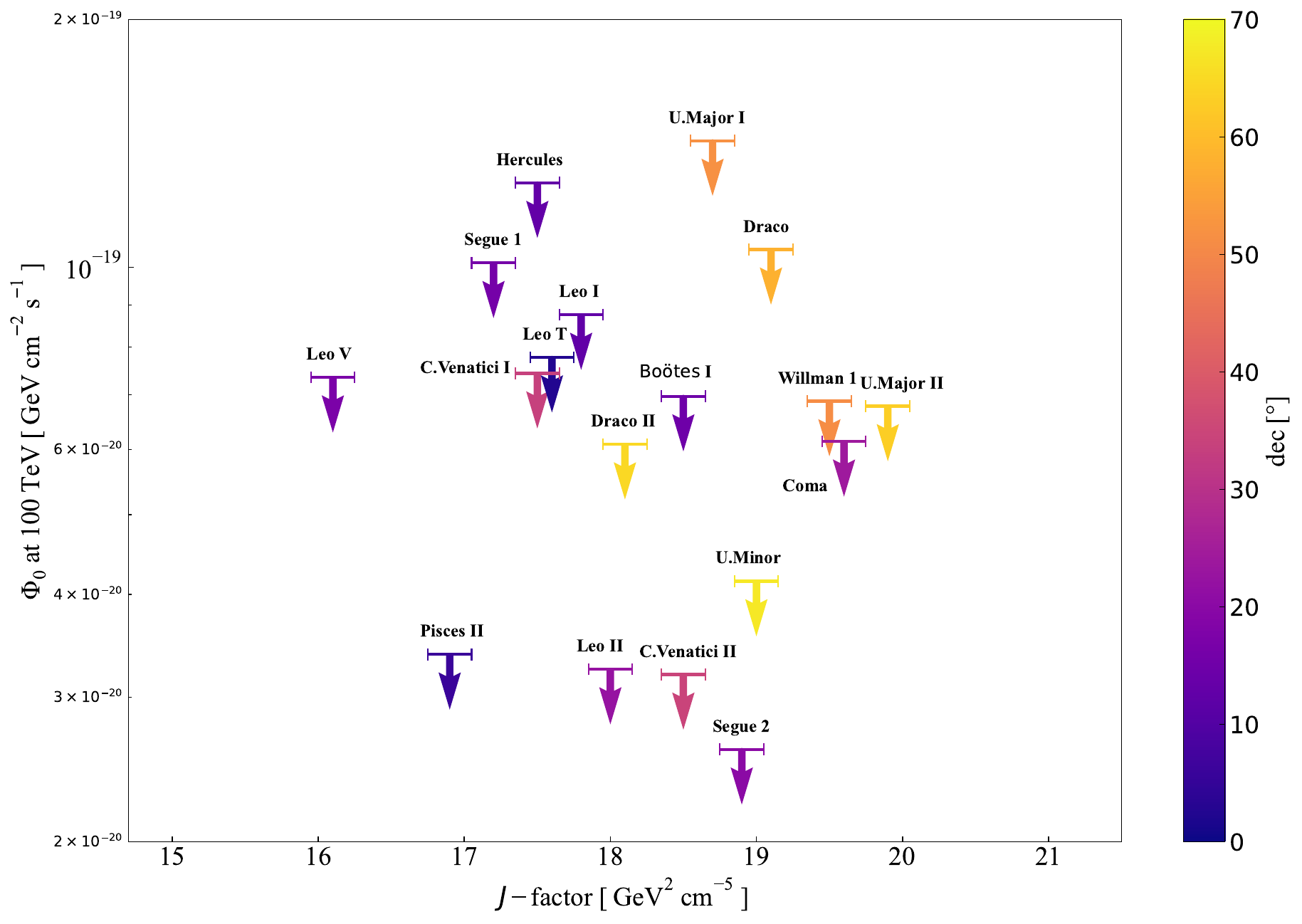}
    \caption{The $x$ axis is J-factors from Table 1, while the $y$ axis is the neutrino flux upper limits at 100 TeV for the 18 dSphs derived from the analysis of IceCube data. The color represents the declination information of the targets. IceCube has a better sensitivity towards the low declination direction. Notice that the horizontal line on each point is not an error bar of J-factor.}
    \label{fig:phi0j}
\end{figure}

\section{results and discussion}

\begin{figure*}[t]
    \centering
    \includegraphics[width=0.9\textwidth]{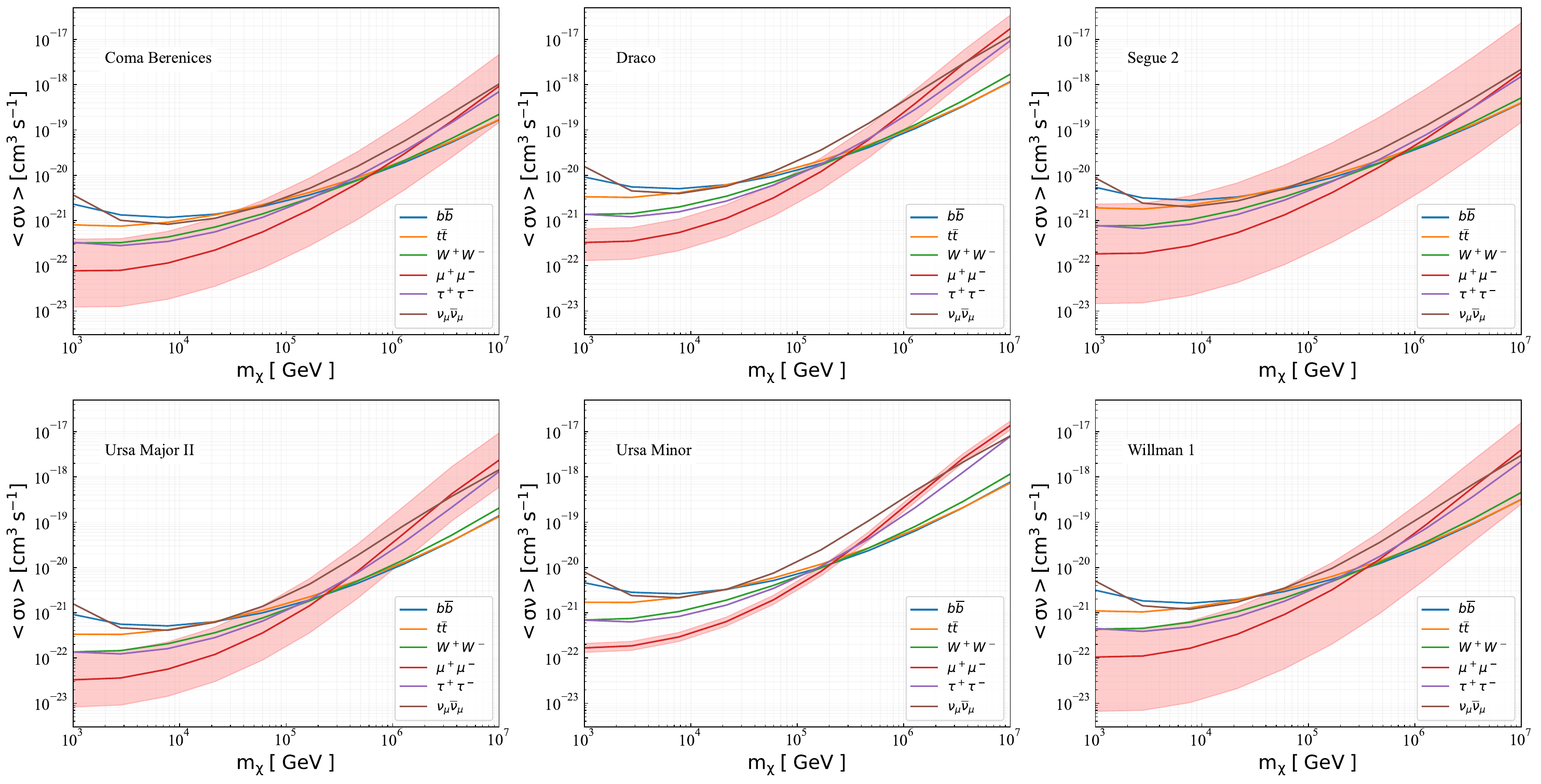}
    \caption{Upper limits on the velocity-averaged annihilation cross section $\left<\sigma v\right>$ derived from single source analysis. Only the 6 sources offering the strongest constraints are present. The shadow band shows the influence by the J-factor uncertainty, which is only drawn on the $\mu^+\mu^-$ channel since it does not change with the annihilation channels.}
    \label{fig:6dsph}
\end{figure*}

\begin{figure}
    \centering
    \includegraphics[width=0.35\textwidth]{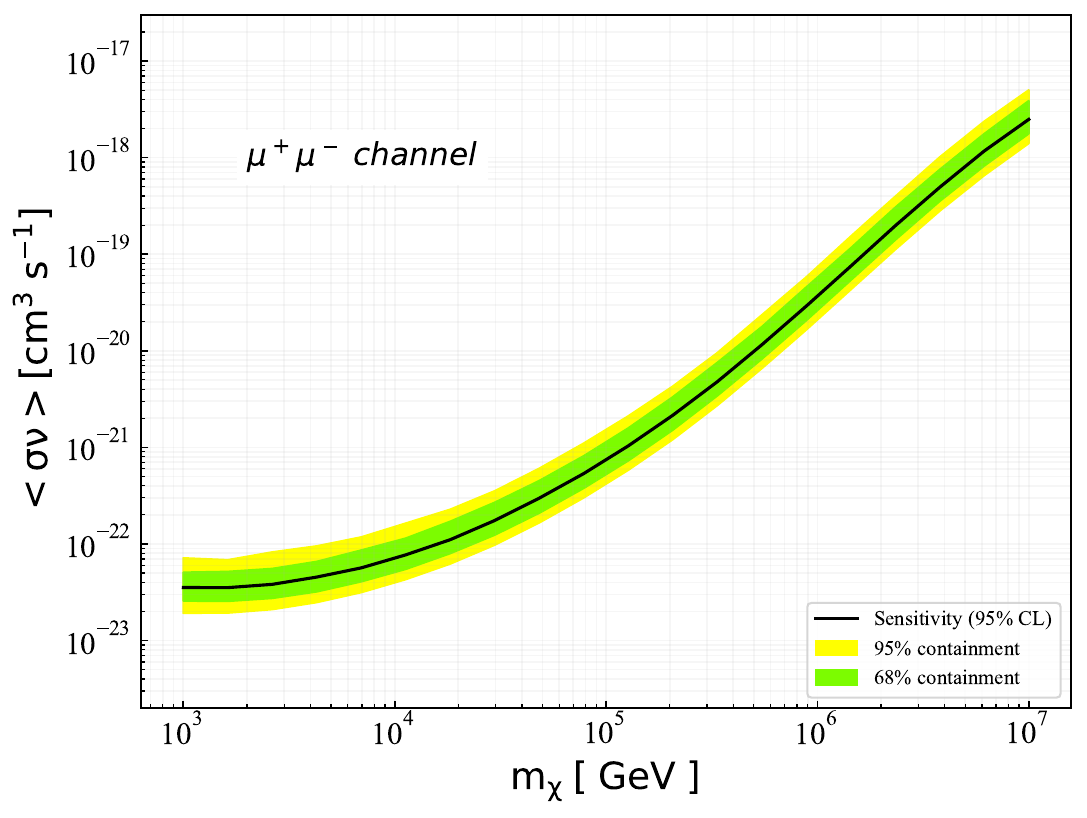}
    \caption{The shaded bands represent 68\% and 95\% containments of limits obtained by 300 blank-sky simulations. The dSph Ursa Major II and the channel $\chi\chi\rightarrow \mu^{+}\mu^{-}$ are chosen for the presentation.}
    \label{fig:mc}
\end{figure}

Using the method described in Sec.~\ref{sec:data-analysis}, we first search for neutrino signals from the directions of the 18 dSphs using a power-law spectrum of $E^{-2.3}$. The results are listed in Table~\ref{tab:tab2}. It can be seen that we do not find any significant signal. The maximum TS value obtained in our analysis is only 2.6 and is from the dSph Hercules. Since no signal is found, we place upper limits on the neutrino fluxes at 100 TeV for these sources, which are listed in the last column of the table. We further in Fig.~\ref{fig:phi0j} present the relationship between the derived flux ULs and J-factors of these dSphs. If there are sub-threshold neutrino emissions, there may be a possible correlation between the flux ULs and the J-factors (because even a weak excess will more-or-less raise the upper limits), but we do not find a significant correlation, further supporting that no underlying signal has been found.
From this figure we can also directly estimate which dSphs are likely to give the strongest limits on the DM parameter.
According to Eq.~(\ref{eq:flux}), sources with larger J-factors and smaller $\Phi_0$ are expected to give the strongest limits. These sources correspond to the lower right corner of Fig.~\ref{fig:phi0j}.

Next, we explore the neutrino signals from dark matter annihilation. We use the spectrum of Eq.~(\ref{eq:flux}) to perform the searches. We scan a series of DM masses from 1 to $10^4$ TeV for different annihilation channels. The DM analyses performed on the 18 dSphs also do not show any evidence for neutrinos emitted from these targets. Consequently, we derive the upper limits at the 95\% CL on the velocity-averaged annihilation cross-section for different DM masses and six channels. The upper limits are shown in Fig.~\ref{fig:6dsph}. We have chosen to show only the 6 sources that offer the strongest constraints, namely Ursa Major II, Coma Berenices, Willman 1, Segue 2, Ursa Minor and Draco. The IceCube observations of dSphs can constrain the $\left<\sigma v\right>$ to a level of $10^{-22}-10^{-21}\,{\rm cm^3s^{-1}}$ for the DM masses of 1-10 TeV.

In Fig.~\ref{fig:mc} we show the 68\% and 95\% containment bands derived from 300 blank-sky simulations. We choose the dSph Ursa Major II and the channel $\chi\chi\rightarrow\mu^+\mu^-$ for the presentation. In each simulation, a random R.A. is assigned and the Dec. is fixed to the one of Ursa Major II. With this new (R.A., Dec.) pair, we perform the analysis with the same procedure as above to derive upper limits. The containment bands represent the expected sensitivity and statistical variation of a background-only analysis of IceCube data.

\begin{figure*}[t]
    \centering
    \includegraphics[width=0.9\textwidth]{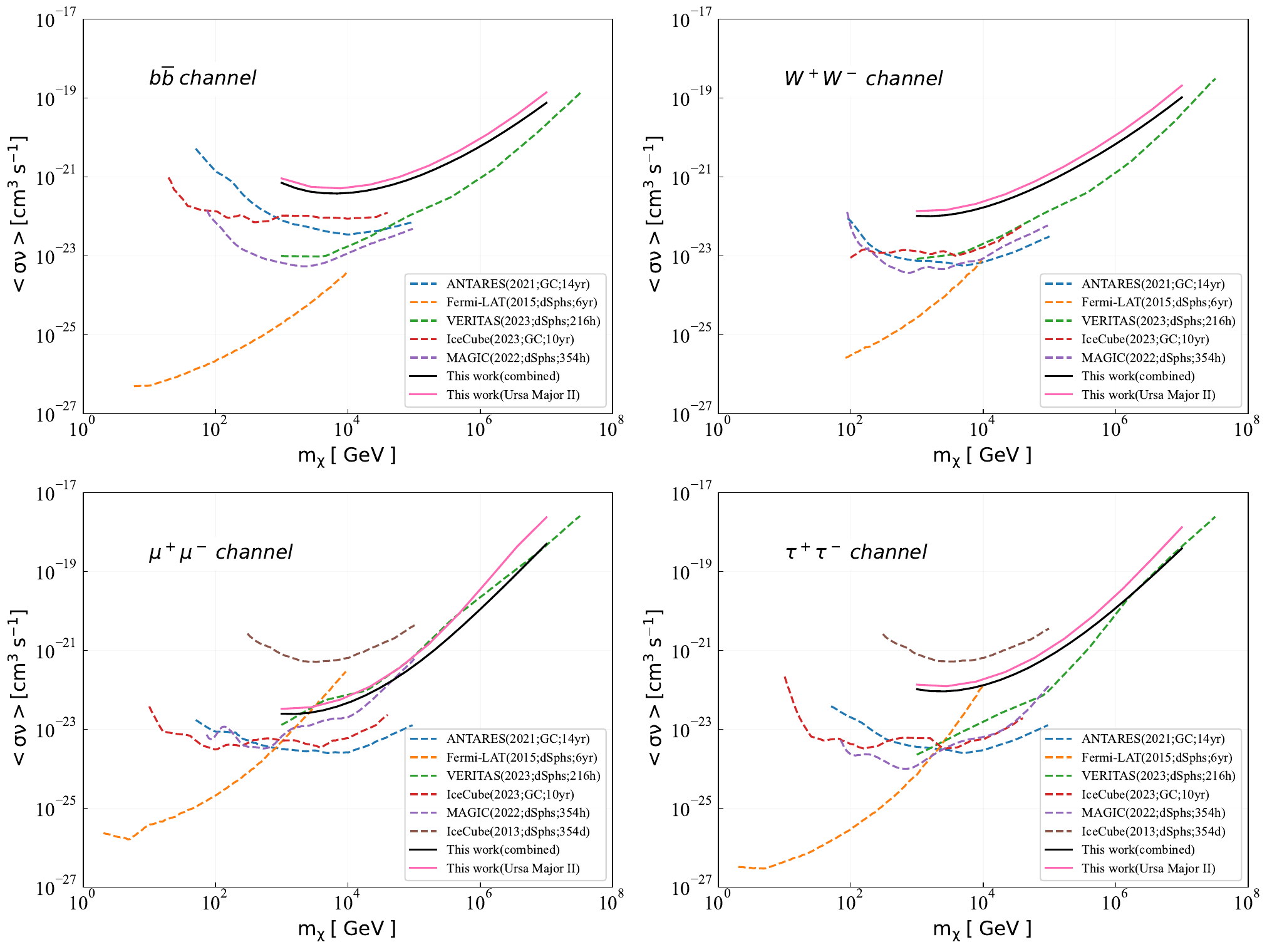}
    \caption{Limits on the velocity-averaged cross-section for the $\chi\chi\rightarrow b\bar{b},W^+W^-,\mu^+\mu^-,\tau^+\tau^-$ channels compared to previous results from IceCube \cite{2013PhRvD..88l2001A,2023arXiv230313663T}, ANTARES \cite{2021JInst..16C9006G} and the gamma-ray telescopes Fermi-LAT \cite{2015PhRvL.115w1301A}, MAGIC \cite{2022PDU....3500912A} and VERITAS \cite{2023ApJ...945..101A}.}
    \label{fig:combinedAna}
\end{figure*}

Since the single-source analysis does not find a signal, we attempt to use an analysis stacking all 18 dSphs together to enhance the search sensitivity. In such a case, the signal PDF is composed of the contributions from all sources (indexed by $j$):
\begin{equation}
    S_i=\frac{\sum_j n_jS_{ij}}{\sum_j n_j}
\end{equation}
where $S_{ij}$ is the signal PDF of a single source (i.e., Eq.~(\ref{eq:si2})) and $n_j$ is the expected neutrino counts from the source $j$, which is given by
\begin{equation}
\begin{aligned}
    n_j(\delta_j,&J_{{\rm ann},j},m_\chi,\left<\sigma v\right>)= \\
    &t\times \int A_{\rm eff}(E_\nu,\delta_j)\Phi_\nu(E_\nu;m_\chi,\left<\sigma v\right>,J_{{\rm ann},j})dE_\nu
\end{aligned}
\end{equation}
and relies on the J-factor ($J_{{\rm ann},j}$) and declination ($\delta_j$) of the $j$-th source. Note that the above two equations should have an additional index $k$ for the 10 data samples, which has been ignored for simplicity. For the background PDF, it has the same form as Eqs.~(\ref{eq:bi1}) and (\ref{eq:bi2}), but the ROI now for the joint analysis is a concatenation of the 18 circular ROIs and the $N_k$ is the total number of events of the new ROI.

The stacking analysis again does not detect a significant neutrino signal. Therefore, we place upper limits on the annihilation cross section based on the stacking analysis, as shown in Fig.~\ref{fig:combinedAna} (black lines). The stacking analysis improves the strongest limits from the single-source analysis (pink line) by a factor of $\sim$2.

Finally, we compare our results with those reported in the literature. It can be seen that at the overlapping energies, the constraints are largely improved compared to the results by \cite{IceCube:2013bas} which is based on the $\sim$1 year of IceCube observation of dSphs in its 59-string configuration. However, the constraints we obtain are weaker than the ones derived from the ANTARES and IceCube observations of the Galactic center \cite{2021JInst..16C9006G,2023arXiv230313663T}, indicating that the strength of the limits is mainly influenced by the J-factor (the GC region has a J-factor of $10^{22}-10^{23}\,\rm GeV^2cm^{-5}$ \cite{Fermi-LAT:2013thd}). For DM searches based on GeV-TeV gamma-ray observations, a key advantage of dSph over the Galactic center is that it has a cleaner astrophysical background \cite{2010PhRvD..81h3506S}. However, for neutrino observations, the galactic center region, although having a more complex astrophysical environment, is also background free in the TeV-PeV energy band, so the magnitude of the J-factor determines the sensitivity of the search. 
At the energies $>10^5\,{\rm GeV}$, the only limits we can find in the literature are from the 216-hour VERITAS observations of dSphs \cite{2023ApJ...945..101A}. Compared to the VERITAS ones, our constraints are comparable if DM particles annihilate through $\mu^+\mu^-$ channel.
For other channels, our constraints are weaker. 
Nevertheless, our results support these previous constraints and are good complements to them.

\section{Summary}
According to mass-to-light ratio measurements, dSphs are rich in DM particles which can produce signals such as gamma-ray photons and neutrinos through annihilation, so dSphs are promising targets for dark matter searches. However, a systematic search for neutrino signals generated by DM annihilation from dSphs is still lacking (see however \cite{2010PhRvD..81h3506S,IceCube:2013bas,2015ICRC...34.1215D}). In this work, we search for neutrino signals from 18 dSphs using 10 years of publicly available muon-track data of the IceCube neutrino observatory. Taking into account the magnitude of the J-factor and the declination at which the source is located, Ursa Major II gives the best result among all 18 sources for single-source analysis. To further improve the sensitivity, we also stack all dSphs together to perform a joint analysis. However, we do not find any significant neutrino emission signal in either the single-source or stacking analysis. Based on such null results, we derive constraints on the annihilation cross section of DM particles. Compared to the existing results in the literature, our constraints are comparable to the ones from the VERITAS observations of dSphs for the $\mu^+\mu^-$ channel and weaker than those given by the ANTARES and IceCube observations of the Galactic center regions at the overlapping energies.

The results of the constraints depend on the accuracy of the J-factors of the dSphs, and there exist uncertainties in the current J-factor measurements given by different groups \cite{2015MNRAS.453..849B,2015ApJ...801...74G,2016PhRvD..93j3512E,Ichikawa:2016nbi,Grand:2020bhk}. 
It also should be noted that most J-factors are calculated assuming spherical symmetry of the dark halos, which may be an oversimplification. Considering the flattening of dark halos will lead to J-factor values different by tens of percent \cite{2016PhRvD..94f3521S}. Upcoming deep imaging surveys such as Vera C. Rubin Observatory Legacy Survey of Space and Time (LSST) \cite{2009arXiv0912.0201L} are hopeful to provide more accurate J-factor measurements, as well as discover more dSphs. The next generation of neutrino detector IceCube-Gen2 \cite{IceCube-Gen2:2020qha} will also improve the sensitivity of neutrino detection. All these will further enhance our capability to search for UHDM.

\begin{acknowledgments}
This work is supported by the National Key Research and Development Program of China (No. 2022YFF0503304), and the Guangxi Science and Technology Program (No. GuiKeAD21220037).
\end{acknowledgments}

\bibliographystyle{apsrev4-1-lyf}
\bibliography{export-bibtex}

\end{document}